\def\be{\begin{equation}}
\def\ee{\end{equation}}
\def\ba{\begin{array}}
\def\ea{\end{array}}

\documentclass[aps,amsmath,amssymb,amsfonts,showpacs,showkeys,pra]{revtex4}
\usepackage{graphicx}
\usepackage{tikz}
\usetikzlibrary{arrows}
\usetikzlibrary{calc,positioning}
\usepackage{amsmath,amsthm}
\def\qed{\leavevmode\unskip\penalty9999 \hbox{}\nobreak\hfill
     \quad\hbox{\leavevmode  \hbox to.77778em{%
               \hfil\vrule   \vbox to.675em%
               {\hrule width.6em\vfil\hrule}\vrule\hfil}}
     \par\vskip3pt}

\begin{document}
\title{Separability criteria via some classes of measurements}
\author{Lu Liu$^{1}$}
\author{Ting Gao$^{1}$}
\email{gaoting@hebtu.edu.cn}
\author{Fengli Yan$^{2}$}
\email{flyan@hebtu.edu.cn}

\affiliation{$^{1}$College of Mathematics and Information Science, Hebei
Normal University, Shijiazhuang 050024, China, $^{2}$College of Physics Science and Information Engineering, Hebei
Normal University, Shijiazhuang 050024, China}

\begin{abstract}

Mutually unbiased bases (MUBs) and symmetric informationally complete (SIC) positive operator-valued measurements (POVMs) are two related topics in quantum information theory. They are generalized to mutually unbiased measurements (MUMs) and general symmetric informationally complete (GSIC) measurements, respectively, that are both not necessarily rank 1. We study the quantum separability problem by using these measurements and present separability criteria for bipartite systems with arbitrary dimensions and multipartite systems of multi-level subsystems. These criteria are proved to be more effective than previous criteria especially when the dimensions of the subsystems are different. Furthermore, full quantum state tomography is not needed when these criteria are implemented in experiment.

\end{abstract}



\maketitle

\section{Introduction }

The concept of entanglement plays a central role in quantum physics and quantum information science, which has been investigated rapidly in recent years \cite{Horodecki09,Nielsen,Guhne09}.
It has numerous applications ranging from quantum cryptography \cite{PRL67.661,77.2816,r1,r2,GYJPA05,r4,r5}, quantum teleportation \cite{PRL70.1895,GYEPL08} to dense coding \cite{PRL69.2881}, and other quantum information processing \cite{r3,r4,r5,r6,r7,r8,r9,r10}.
One of the important tasks of the theory of quantum entanglement is to characterize entanglement.
Although many important results have been obtained for bipartite systems \cite{PRL77.1413,QIC3.193,PRA59.4206,PRL99.130504} and multipartite systems \cite{PRL.113.100501,PRA.82.062113,EPJD.61.765,NJP.12.053002,QIC.10.829,PRL.112.180501,EPL.104.20007,PRA.93.042310,PRA.86.062323}, a general theory remains elusive because of the complexity of entanglement.
Recently, because of its significant role in quantum information processing, much effort has been devoted to investigate various measurements that can be used for the detection of entanglement of unknown quantum states.

Mutually unbiased bases (MUBs) represent maximally non-commutative measurements.
They were used for detecting entangled states in two-qudit quantum systems \cite{PRA.86.022311}. However, when $d$ is not a prime power, the maximum number $N(d)$ of MUBs remains open \cite{1409.3386[11]}, which makes the criterion becomes less effective.

Mutually unbiased measurements (MUMs) were generalized from MUBs \cite{NJP.16.053038} and include the complete set of MUBs as a special case.
The existence of MUMs does not depend on the dimension of the system, and a complete set of MUMs were constructed for arbitrary finite dimensional Hilbert space in Ref.\cite{NJP.16.053038}. They were used to construct separability criteria in bipartite finite dimensional systems \cite{PRA.89.064302,1407.7333}.

The notion of symmetric informationally complete (SIC) positive operator-valued measurements (POVMs) is another related topic in quantum information theory.
It was generalized to general symmetric informationally complete (GSIC) measurements, of which the complete sets were constructed in all finite dimensions without requirement that the rank of each measurement operator is one \cite{1305.6545}. A separability criterion for $d$-dimensional bipartite systems via GSIC-POVMs was presented in Ref.\cite{1406.7820}.

In Ref.\cite{PRA.91.012326}, three separability criteria were proposed based on $\rho-\rho^{A}\bigotimes\rho^{B}$, where $\rho$ is a bipartite density matrix in $\mathbb{C}^{d}\bigotimes \mathbb{C}^{d}$ and $\rho^{A}(\rho^{B})$ is the reduced density matrix of the first (second) subsystem.

Most of the criteria using MUBs, MUMs, GSIC-POVMs mentioned above are for $d$-dimensional bipartite systems, of which the subsystems should be with the same dimension. We obtained separability criteria on arbitrary high-dimensional bipartite systems of a $d_1$-dimensional subsystem and a $d_2$-dimensional subsystem, and multipartite systems of multi-level subsystems \cite{SR,1512.02853} such that the criteria for $d$-dimensional bipartite systems in Ref.\cite{1407.7333,PRA.91.012326} are the special cases of ours. However, the criteria in \cite{SR,1512.02853} are not efficient enough because the bounds are not tight. Thus, how to use the sets of these measurements to detect entanglement more efficiently still need to be considered.

In this paper, we study the separability problem via sets of MUMs and propose more effective separability criteria for $\mathbb{C}^{d} \otimes \mathbb{C}^{d'}$ systems. Without difficulty, our method can be used to construct separability criteria via MUBs and GSIC-POVMs in $\mathbb{C}^{d} \otimes \mathbb{C}^{d'}$ systems or high-dimensional multipartite systems.

\section{MAIN RESULTS}

\noindent

For the bipartite system of subsystems with different dimensions, the complete sets of MUMs cannot be used to detect the separability of quantum states in Ref.\cite{SR}. This problem will be discussed and we obtain the following conclusions.

\vspace{0.2cm}{{\slshape Theorem 1.} Let $\rho$ be a density matrix in $\mathbb{C}^{d} \otimes \mathbb{C}^{d'}$, without loss of generality let $d<d'$, $d'=sd+r_{1}$. $\{\mathcal{P}^{(b)}\}_{b=1}^{M}$ ($\{\mathcal{Q}^{(b)}\}_{b=1}^{M'}$) are any sets of $M$ ($M'$) $MUMs$ on $\mathbb{C}^{d}$ ($\mathbb{C}^{d'}$)  with efficiency parameter $\kappa_{1}$ ($\kappa_{2}$), where $\mathcal{P}^{(b)}=\{P_{n}^{(b)}\}_{n=1}^{d}$, $\mathcal{Q}^{(b)}=\{Q_{n'}^{(b)}\}_{n'=1}^{d'}$, and $M'=tM+r_{2}$.
Define
\begin{equation}
\begin{array}{ll}
J(\rho)=\max\limits_{\begin{subarray}{c} \{Q_{n_{p}}^{(b_{q})}\}\subseteq\mathcal{Q}^{(b)} 
 \end{subarray} }
\sum\limits_{b=1}^{M}\sum\limits_{q=1}^{t}\sum\limits_{n=1}^{d}\sum\limits_{p=1}^{s}
\mathrm{Tr}(P_{n}^{(b)}\otimes Q_{n_{p}}^{(b_{q})}\rho).
\end{array}
\end{equation}

If $\rho$ is separable, then
\begin{equation}
J(\rho)\leq\frac{ts}{2}\left(\frac{M-1}{d}+\kappa_{1}\right)+\frac{1}{2}\left(\frac{M'-1}{d'}+\kappa_{2}\right).\label{th1}
\end{equation}

\vspace{0.2cm}

{\slshape Proof.}~It's only needed to consider a pure separable state $\rho=|\phi\rangle\langle\phi|\otimes|\psi\rangle\langle\psi|$,
since $\sum\limits_{b=1}^{M}\sum\limits_{q=1}^{t}\sum\limits_{n=1}^{d}\sum\limits_{p=1}^{s}
\mathrm{Tr}(P_{n}^{(b)}\otimes Q_{n_{p}}^{(b_{q})}\rho)$ is a linear function of $\rho$. We have
\begin{equation}
\begin{array}{ll}
&~~~~~\sum\limits_{b=1}^{M}\sum\limits_{q=1}^{t}\sum\limits_{n=1}^{d}\sum\limits_{p=1}^{s}
               \mathrm{Tr}(P_{n}^{(b)}\otimes Q_{n_{p}}^{(b_{q})}\rho)\\
&\leq 
\frac{1}{2}\sum\limits_{b=1}^{M}\sum\limits_{q=1}^{t}\sum\limits_{n=1}^{d}\sum\limits_{p=1}^{s}
      [\textrm{Tr}(P_{n}^{(b)}|\phi\rangle\langle\phi|)]^{2}
  +\frac{1}{2}\sum\limits_{b=1}^{tM}\sum\limits_{n=1}^{sd}
      [\textrm{Tr}(Q_{n}^{(b)}|\psi\rangle\langle\psi|)]^{2}\\
&\leq \frac{1}{2}\sum\limits_{b=1}^{M}\sum\limits_{q=1}^{t}\sum\limits_{n=1}^{d}\sum\limits_{p=1}^{s}
      [\textrm{Tr}(P_{n}^{(b)}|\phi\rangle\langle\phi|)]^{2}
  +\frac{1}{2}\sum\limits_{b=1}^{M'}\sum\limits_{n=1}^{d'}
      [\textrm{Tr}(Q_{n}^{(b)}|\psi\rangle\langle\psi|)]^{2}\\
&\leq \frac{ts}{2}(\frac{M-1}{d}+\kappa_{1})+\frac{1}{2}(\frac{M'-1}{d'}+\kappa_{2}),
\end{array}
\end{equation}
where the inequality \cite{1407.7333}
 \begin{equation}
\sum\limits_{b=1}^{M}\sum\limits_{n=1}^{d}\textrm{Tr}(P_{n}^{(b)}\rho)^{2}
\leq\frac{M-1}{d}+\frac{1-\kappa+(\kappa d-1)\textrm{Tr}(\rho^{2})}{d-1} \label{M-1}
\end{equation}
is used.
This completes the proof. \hfill $\square$

Theorem 1 is more effective than Theorem 2 in Ref.\cite{SR} as long as the two subsystems have different dimensions, which can be used for a wider range of application. In Theorem 1, $M$ and $M'$ can be different, while in  Theorem 2 in Ref.\cite{SR}, they are equal. When $M=M'$, by Theorem 1, we obtain that if the left hand side of (\ref{th1}) is larger than $\frac{s}{2}(\frac{M-1}{d}+\kappa_{1})+\frac{1}{2}(\frac{M-1}{d'}+\kappa_{2})$,
then $\rho$ is entangled, while by Theorem 2 in Ref.\cite{SR}, one can derive that if  the left hand side of (\ref{th1}) is larger than $\frac{s}{2}[(\frac{M-1}{d}+\kappa_{1})+(\frac{M-1}{d'}+\kappa_{2})]$, then $\rho$ is entangled. That is, Theorem 1 detects states $\rho$,   for $\begin{array}{ll}
J(\rho)=\max\limits_{\begin{subarray}{c} \{Q_{n_{p}}^{(b_{q})}\}\subseteq\mathcal{Q}^{(b)} 
 \end{subarray} }
\sum\limits_{b=1}^{M}\sum\limits_{q=1}^{t}\sum\limits_{n=1}^{d}\sum\limits_{p=1}^{s}
\mathrm{Tr}(P_{n}^{(b)}\otimes Q_{n_{p}}^{(b_{q})}\rho)
\end{array} > \frac{s}{2}(\frac{M-1}{d}+\kappa_{1})+\frac{1}{2}(\frac{M-1}{d'}+\kappa_{2})$,  as entangled, whereas Theorem 2 in Ref.\cite{SR} detects them only for $\begin{array}{ll}
J(\rho)=\max\limits_{\begin{subarray}{c} \{Q_{n_{p}}^{(b_{q})}\}\subseteq\mathcal{Q}^{(b)} 
 \end{subarray} }
\sum\limits_{b=1}^{M}\sum\limits_{q=1}^{t}\sum\limits_{n=1}^{d}\sum\limits_{p=1}^{s}
\mathrm{Tr}(P_{n}^{(b)}\otimes Q_{n_{p}}^{(b_{q})}\rho)
\end{array} > \frac{s}{2}[(\frac{M-1}{d}+\kappa_{1})+(\frac{M-1}{d'}+\kappa_{2})]$.  Therefore, Theorem 1 is better than Theorem 2 in Ref.\cite{SR}, especially when the difference of the dimensions of two subsystems is very large.
 What's more, no term is needed to be ignored by Theorem 1 when $d'$ is a multiple of $d$ and $M'$ is a multiple of $M$, so that Theorem 1 is much more effective. When $d|d'$ and $(d+1)|(d'+1)$, we can detect $\mathbb{C}^{d} \otimes \mathbb{C}^{d'}$ entangled states using complete sets of MUMs by Theorem 1.

With the help the Cauchy-Schwarz inequality, we can obtain stronger bound than that in Theorem 1.

\vspace{0.2cm}{\slshape Theorem 2.} Let $\rho$ be a density matrix in $\mathbb{C}^{d} \otimes \mathbb{C}^{d'}$, without loss of generality let $d<d'$, $d'=sd+r_{1}$, and $\{\mathcal{P}^{(b)}\}_{b=1}^{M}$ and
$\{\mathcal{Q}^{(b)}\}_{b=1}^{M'}$ be any two sets of $M$ and $M'$ $MUMs$ on $\mathbb{C}^{d}$ and $\mathbb{C}^{d'}$ with efficiency parameters $\kappa_{1}$, $\kappa_{2}$, respectively, where $\mathcal{P}^{(b)}=\{P_{n}^{(b)}\}_{n=1}^{d}$, and $\mathcal{Q}^{(b)}=\{Q_{n'}^{(b)}\}_{n'=1}^{d'}$, $M'=tM+r_{2}$.
If $\rho$ is separable, then it satisfies the following inequality
\begin{equation}
J(\rho)\leq\sqrt{ts(\frac{M-1}{d}+\kappa_{1})}\sqrt{\frac{M'-1}{d'}+\kappa_{2}}.\label{th2}
\end{equation}
Here $J(\rho)$ is defined the same as in Theorem 1.

\vspace{0.2cm}
{\slshape Proof.}~ For a pure separable state $\rho=|\phi\rangle\langle\phi|\otimes|\psi\rangle\langle\psi|$,
we get
\begin{equation}
\begin{array}{ll}
&~~~~~\sum\limits_{b=1}^{M}\sum\limits_{q=1}^{t}\sum\limits_{n=1}^{d}\sum\limits_{p=1}^{s}
               \mathrm{Tr}(P_{n}^{(b)}\otimes Q_{n_{p}}^{(b_{q})}\rho)\\
&=\sum\limits_{b=1}^{M}\sum\limits_{q=1}^{t}\sum\limits_{n=1}^{d}\sum\limits_{p=1}^{s}
   \textrm{Tr}(P_{n}^{(b)}|\phi\rangle\langle\phi|) \textrm{Tr}(Q_{n_{p}}^{(b_{q})}|\psi\rangle\langle\psi|)\\
&\leq \sqrt{\sum\limits_{b=1}^{M}\sum\limits_{q=1}^{t}\sum\limits_{n=1}^{d}\sum\limits_{p=1}^{s}
            [\textrm{Tr}(P_{n}^{(b)}|\phi\rangle\langle\phi|)]^{2}}
      \sqrt{\sum\limits_{b=1}^{M}\sum\limits_{q=1}^{t}\sum\limits_{n=1}^{d}\sum\limits_{p=1}^{s}
            [\textrm{Tr}(Q_{n_{p}}^{(b_{q})}|\psi\rangle\langle\psi|))]^{2}}\\
&\leq
\sqrt{ts(\frac{M-1}{d}+\kappa_{1})}\sqrt{\frac{M'-1}{d'}+\kappa_{2}},
\end{array}
\end{equation}
where the Cauchy-Schwarz inequality and the inequality (\ref{M-1}) are used. It is easily to see that $\sum\limits_{b=1}^{M}\sum\limits_{q=1}^{t}\sum\limits_{n=1}^{d}\sum\limits_{p=1}^{s}
\mathrm{Tr}(P_{n}^{(b)}\otimes Q_{n_{p}}^{(b_{q})}\rho)$ is a linear function of $\rho$, so the inequality (\ref{th2}) holds for separable mixed states.
This completes the proof. \hfill $\square$

The bound in Theorem 2 is lower than that in Theorem 1 since  $\sqrt{ts(\frac{M-1}{d}+\kappa_{1})}\sqrt{\frac{M'-1}{d'}+\kappa_{2}}\leq \frac{1}{2}\big(ts(\frac{M-1}{d}+\kappa_{1})+\frac{M'-1}{d'}+\kappa_{2}\big)$.

With the same method, we can obtain separability criteria using MUBs and GSIC-POVMs.

\vspace{0.2cm}{\slshape Theorem 2~$'$(MUBs).} Let $\rho$ be a density matrix in $\mathbb{C}^{d} \otimes \mathbb{C}^{d'}$, without loss of generality let $d<d'$, $d'=sd+r_{1}$, and $\mathcal{D}_{1}=\{\mathcal{B}_{1,1},\mathcal{B}_{1,2},\cdots,\mathcal{B}_{1,M}\}$,
~$\mathcal{D}_{2}=\{\mathcal{B}_{2,1},\mathcal{B}_{2,2},\cdots,\mathcal{B}_{2,M'}\}$ be two sets of MUBs on~$\mathbb{C}^{d_{1}}$,~$\mathbb{C}^{d_{2}}$, respectively, where ~$\mathcal{B}_{1,k}=\{|\sigma_{i}^{k}\rangle\}_{i=1}^{d}$,~$\mathcal{B}_{2,k}=\{|\tau_{j}^{k}\rangle\}_{j=1}^{d'}$, and~$M'=tM+r_{2}$.
Define
\begin{equation}
J(\rho)=\max\limits_{ \{|\tau_{i_{u}}^{k_{v}}\rangle\}\subseteq\mathcal{B}_{2,k}}
     \sum\limits_{k=1}^{M}\sum\limits_{i=1}^{d}\sum\limits_{u=1}^{s}\sum\limits_{v=1}^{t}
        \langle \sigma_{i}^{k}|\otimes\langle \tau_{i_{u}}^{k_{v}}|\rho |\sigma_{i}^{k}\rangle\otimes|\tau_{i_{u}}^{k_{v}}\rangle,
\end{equation}
If $\rho$ is separable, then
\begin{equation}
J(\rho)\leq \sqrt{ts(1+\frac{M-1}{d})}\sqrt{1+\frac{M'-1}{d'}}.
\end{equation}

{\slshape Proof.}~For separable state  $\rho=\sum\limits_j p_j \rho_j^1\otimes \rho_j^2$, where $\rho_j^1$ and $\rho_j^2$ are pure states in $\mathbb{C}^{d}$ and $\mathbb{C}^{d'}$, respectively, there is
\begin{equation}
\begin{array}{ll}
&\sum\limits_{k=1}^{M}\sum\limits_{i=1}^{d}\sum\limits_{u=1}^{s}\sum\limits_{v=1}^{t}
        \langle \sigma_{i}^{k}|\otimes\langle \tau_{i_{u}}^{k_{v}}|\rho |\sigma_{i}^{k}\rangle\otimes|\tau_{i_{u}}^{k_{v}}\rangle\\
= & \sum\limits_j p_j\sum\limits_{k=1}^{M}\sum\limits_{i=1}^{d}\sum\limits_{u=1}^{s}\sum\limits_{v=1}^{t}
   \langle \sigma_{i}^{k}|\rho_j^1 | \sigma_{i}^{k}\rangle
    \langle \tau_{i_{u}}^{k_{v}}|\rho_j^2|\tau_{i_{u}}^{k_{v}}\rangle\\
\leq & \sum\limits_j p_j\sqrt{\sum\limits_{k=1}^{M}\sum\limits_{i=1}^{d}\sum\limits_{u=1}^{s}\sum\limits_{v=1}^{t}
      \langle \sigma_{i}^{k}|\rho_j^1 | \sigma_{i}^{k}\rangle^2}
      \sqrt{\sum\limits_{k=1}^{M}\sum\limits_{i=1}^{d}\sum\limits_{u=1}^{s}\sum\limits_{v=1}^{t}
      \langle \tau_{i_{u}}^{k_{v}}|\rho_j^2|\tau_{i_{u}}^{k_{v}}\rangle^2}\\
\leq & \sum\limits_j p_j\sqrt{\sum\limits_{k=1}^{M}\sum\limits_{i=1}^{d}\sum\limits_{u=1}^{s}\sum\limits_{v=1}^{t}
                  \langle \sigma_{i}^{k}|\rho_j^1 | \sigma_{i}^{k}\rangle^2}
     \sqrt{\sum\limits_{k=1}^{M'}\sum\limits_{j=1}^{d'}
                  \langle \tau_{i_{u}}^{k_{v}}|\rho_j^2|\tau_{i_{u}}^{k_{v}}\rangle^2}\\
\leq & \sqrt{ts(1+\frac{M-1}{d})}\sqrt{1+\frac{M'-1}{d'}},
\end{array}
\end{equation}
where the inequality~\cite{PRA.79.022104}
\begin{equation}\label{B}
\sum\limits_{k=1}^{M}\sum\limits_{i=1}^{d}\langle \sigma_{i}^{k}|\rho_j^1 | \sigma_{i}^{k}\rangle^2\leq 1+\frac{M-1}{d}
\end{equation}
is used.
This completes the proof. \hfill $\square$  \\

By an analogous argument as Theorem 2 and using the inequality \cite{PS89.085101}
\begin{equation}\label{SIC}
\sum\limits_{n=1}^{d^{2}}[\textrm{Tr}(P_{n}\rho)]^{2}=\frac{(ad^{3}-1)\textrm{Tr}(\rho^2)+d(1-ad)}{d(d^2-1)},
\end{equation}
 we get the following result.

\vspace{0.2cm}{\slshape Theorem 2~$''$(GSIC-POVMs).} Let $\rho$ be a density matrix in $\mathbb{C}^{d} \otimes \mathbb{C}^{d'}$, without loss of generality let $d<d'$, $d'^2=sd^2+r$, and ~$\mathcal{P}_{1}$,~$\mathcal{P}_{2}$ are two sets of GSIC-POVMs on ~$\mathbb{C}^{d}$,~$\mathbb{C}^{d'}$ with efficiency parameters ~$a_{1}$,~$a_{2}$, respectively.
Define
\begin{equation}
J(\rho)=\max_{\begin{subarray}{c} \{P_{1,n}\}\subseteq\mathcal{P}_{1}\\
                                    \{P_{2,n_i}\}\subseteq\mathcal{P}_{2} \end{subarray}}
\sum\limits_{n=1}^{d^{2}}\sum\limits_{i=1}^{s}\textrm{Tr}(P_{1,n}\otimes P_{2,n_i}\rho).
\end{equation}
If $\rho$ is separable, then
\begin{equation}
J(\rho)\leq
      \sqrt{\frac{s(a_{1}d^{2}+1)}{d(d+1)}}
      \sqrt{\frac{a_{2}d'^{2}+1}{d'(d'+1)}}.
\end{equation}

Inspired by the separability criteria based on the operators \cite{PRA.91.012326,1512.02853}
\begin{equation}\label{deltarou}
    \Delta\rho=\frac{1}{2^{N-2}}(\mathcal{Q_{II}}-\mathcal{Q_{I}}),
\end{equation}
where $N$ is an even number, $\mathcal{Q_{II}}=\sum_{q\in\mathcal{P_{II}}}q$, $\mathcal{Q_{I}}=\sum_{p\in\mathcal{P_{I}}}p$ and $\mathcal{P_{I}}$ ($\mathcal{P_{II}}$) denotes that both sides of bipartite partition contain odd (even) number of parties, we deduce the next theorem.

\vspace{0.2cm} {\slshape Theorem 3.} Let $\rho$ be a density matrix in $\mathbb{C}^{d} \otimes \mathbb{C}^{d'}$, without loss of generality let $d<d'$, $d'=sd+r_{1}$, and $\{\mathcal{P}^{(b)}\}_{b=1}^{M}$ and
$\{\mathcal{Q}^{(b)}\}_{b=1}^{M'}$ be any two sets of $M$ and $M'$ $MUMs$ on $\mathbb{C}^{d}$ and $\mathbb{C}^{d'}$ with efficiency parameters $\kappa_{1}$, $\kappa_{2}$, respectively, where $\mathcal{P}^{(b)}=\{P_{n}^{(b)}\}_{n=1}^{d}$, and $\mathcal{Q}^{(b)}=\{Q_{n'}^{(b)}\}_{n'=1}^{d'}$, $M'=tM+r_{2}$.
Define
$$S(\rho)=\sum\limits_{b=1}^{M}\sum\limits_{q=1}^{t}\sum\limits_{n=1}^{d}\sum\limits_{p=1}^{s}
\big|\mathrm{Tr}(P_{n}^{(b)}\otimes Q_{n_{p}}^{(b_{q})}(\rho-\rho^{A}\otimes\rho^{B}))|.$$
The following inequality
\begin{equation}
S(\rho)\leq\sqrt{ts\{(\frac{M-1}{d}+\kappa_{1})-\sum\limits_{b=1}^{M}\sum\limits_{n=1}^{d}
                 [\mathrm{Tr}(P_{n}^{(b)}\rho^{A})]^{2}\}}
           \sqrt{\frac{M'-1}{d'}+\kappa_{2}-\sum\limits_{b=1}^{M'}\sum\limits_{n=1}^{d'}
                 [\mathrm{Tr}(Q_{n}^{(b)}\rho^{B})]^{2}}\label{th3}
\end{equation}
holds for separable states $\rho$.

{\slshape Proof.}~
Note that for any separable state $\rho$, $\rho-\rho^{A}\otimes\rho^{B}$ can be written as the form of \cite{PRA.77.060301}
$$
\rho-\rho^{A}\otimes\rho^{B}=\frac{1}{2}\sum_{u,v}p_{u}p_{v}(\rho_{u}^{A}-\rho_{v}^{A})
\otimes(\rho_{u}^{B}-\rho_{v}^{B}),
$$
where $\rho_{u}^{A}$ and $\rho_{u}^{B}$ are the pure states density matrix acting on the first and second subsystem, respectively.
There is
\begin{equation*}
\begin{array}{ll}
&\sum\limits_{b=1}^{M}\sum\limits_{q=1}^{t}\sum\limits_{n=1}^{d}\sum\limits_{p=1}^{s}
              \big|\mathrm{Tr}(P_{n}^{(b)}\otimes Q_{n_{p}}^{(b_{q})})(\rho-\rho^{A}\otimes\rho^{B})\big|\\
\leq&\sum\limits_{b=1}^{M}\sum\limits_{q=1}^{t}\sum\limits_{n=1}^{d}\sum\limits_{p=1}^{s}
              \sum\limits_{u,v}\frac{1}{2}p_{u}p_{v}\big|\textrm{Tr}(P_{n}^{(b)}(\rho_{u}^{A}-\rho_{v}^{A}))\big|
                                                   \big|\textrm{Tr}(Q_{n_{p}}^{(b_{q})}(\rho_{u}^{B}-\rho_{v}^{B}))\big|\\
\leq&\sqrt{\sum\limits_{b=1}^{M}\sum\limits_{q=1}^{t}\sum\limits_{n=1}^{d}\sum\limits_{p=1}^{s}\{\sum\limits_{u}p_{u}
                        [\textrm{Tr}(P_{n}^{(b)}\rho_{u}^{A})]^{2}-[\textrm{Tr}(P_{n}^{(b)}\rho^{A})]^{2}\}}
     \sqrt{\sum\limits_{b=1}^{M'}\sum\limits_{n=1}^{d'}\{\sum\limits_{u}p_{u}
                        [\textrm{Tr}(Q_{n_{p}}^{(b_{q})}\rho_{u}^{B})]^{2}-[\textrm{Tr}(Q_{n_{p}}^{(b_{q})}\rho^{B})]^{2}\}}\\
\leq&\sqrt{ts\{(\frac{M-1}{d}+\kappa_{1})-\sum\limits_{b=1}^{M}\sum\limits_{n=1}^{d}
                 [\mathrm{Tr}(P_{n}^{(b)}\rho^{A})]^{2}\}}
           \sqrt{\frac{M'-1}{d'}+\kappa_{2}-\sum\limits_{b=1}^{M'}\sum\limits_{n=1}^{d'}
                 [\mathrm{Tr}(Q_{n}^{(b)}\rho^{B})]^{2}},
\end{array}
\end{equation*}
as required.
\hfill $\square$

To show that Theorem 3 is stronger than Theorem 1, we only need to prove that the inequality (\ref{th1}) holds if (\ref{th3}) holds. In fact,
inequality(\ref{th3}) implies that
\begin{equation*}
\begin{array}{ll}
            &\sum\limits_{b=1}^{M}\sum\limits_{q=1}^{t}\sum\limits_{n=1}^{d}\sum\limits_{p=1}^{s}
              \mathrm{Tr}(P_{n}^{(b)}\otimes Q_{n_{p}}^{(b_{q})}\rho)\\
       \leq & S(\rho)
             +\sum\limits_{b=1}^{M}\sum\limits_{q=1}^{t}\sum\limits_{n=1}^{d}\sum\limits_{p=1}^{s}
               \mathrm{Tr}[(P_{n}^{(b)}\otimes Q_{n_{p}}^{(b_{q})})(\rho^{A}\otimes\rho^{B})]\\
       \leq & \frac{ts}{2}(\frac{M-1}{d}+\kappa_{1})+\frac{1}{2}(\frac{M'-1}{d'}+\kappa_{2})
             -\frac{1}{2}\sum\limits_{b=1}^{M}\sum\limits_{q=1}^{t}\sum\limits_{n=1}^{d}\sum\limits_{p=1}^{s}
              \{[\mathrm{Tr}P_{n}^{(b)}\rho^{A}]^{2}+[\mathrm{Tr}Q_{n_{p}}^{(b_{q})}\rho^{B}]^{2}
                -2\mathrm{Tr}(P_{n}^{(b)}\otimes Q_{n_{p}}^{(b_{q})}\rho^{A}\otimes\rho^{B})\}\\
       \leq & \frac{ts}{2}(\frac{M-1}{d}+\kappa_{1})+\frac{1}{2}(\frac{M'-1}{d'}+\kappa_{2}).
\end{array}
\end{equation*}
Thus, the inequality (\ref{th1}) holds.

The separability criteria in Ref.\cite{PRA.86.022311,PRA.89.064302,1407.7333,1406.7820} are all only applied for quantum systems of subsystems with the same dimension. For that with different dimensions, we obtained separability criterion in Ref.\cite{SR}, which was discussed less efficient than the criteria in this paper. In brief, the criteria we present here is more efficient and wider range of application.

Noting the significance of the study on multiparty quantum entanglement, especially in higher-dimensional systems, we generalize our criteria to high dimensional multipartite systems.

\vspace{0.2cm} {\slshape Theorem 4.}~Suppose that $\rho$ is a density matrix in $\mathbb{C}^{d_{1}}\otimes\mathbb{C}^{d_{2}}\otimes\cdots\otimes\mathbb{C}^{d_{m}}$, $\mathcal{P}^{(b)}_{k}$ are any sets of $M_{k}$ MUMs on $\mathbb{C}^{d_{k}}$ with the efficiency parameter $\kappa_{k}$, and $d_{k}=s_{k}d+r_{1k}, M_{k}=t_{k}M+r_{2k}$,
where $d=\min \{d_{1},d_{2},\cdots,d_{m}\}$, $M=\min \{M_{1},M_{2},\cdots,M_{m}\}$.
Define
\begin{equation}\label{}
J(\rho)=
\sum\limits_{j=1}^{M}\sum\limits_{i=1}^{d}
\mathrm{Tr}\Big(\big(\otimes^{m}_{k=1}(\sum\limits_{p=1}^{s_{k}}\sum\limits_{q=1}^{t_{k}}P_{k,(i,p)}^{(j,q)})\big) \rho\Big),
\end{equation}
where $P_{k,(i,p)}^{(j,q)}\in \mathcal{P}^{(b)}_{k}$.
For any fully separable state $\rho$, it satisfies the following inequalities:
\begin{equation}\label{th4-1}
J(\rho)\leq \frac{1}{m}\sum\limits_{i=1}^{m}\frac{\prod_{k=1}^{m} s_{k}t_{k}}{s_{i}t_{i}}\Big(\frac{M_{i}-1}{d_{i}}+\kappa_{i}\Big),
\end{equation}
\begin{equation}\label{th4-2}
J(\rho)\leq \min_{1\leq i\neq j\leq m}\prod_{k=1}^{m} s_{k}t_{k}\sqrt{\frac{1}{s_{i}t_{i}}(\frac{M_{i}-1}{d_{i}}+\kappa_{i})}\sqrt{\frac{1}{s_{j}t_{j}}(\frac{M_{j}-1}{d_{j}}+\kappa_{j})}.
\end{equation}

{\slshape Proof.}~
~Let $\rho=\sum\limits_{l}p_{l}\rho_{l}=\sum\limits_{l}p_{l}\rho_{1l}\otimes\rho_{2l}\cdots \otimes \rho_{ml}$ with $\sum\limits_{l}p_{l}=1$, be a fully separable density matrix, where $\rho_{kl}$ are pure states in $\mathbb{C}^{d_{k}}$. Note that
 $0\leq \textrm{Tr}(P_{k,(i,p_{k})}^{(j,q_{k})}\rho_{kl})\leq 1$, by using Lemma 1 of Ref.\cite{SR}, we have
\begin{equation}
\begin{array}{rl}
J(\rho_l)=&\sum\limits_{j=1}^{M}\sum\limits_{i=1}^{d}
 \sum\limits_{p_{1}=1}^{s_{1}}\sum\limits_{q_{1}=1}^{t_{1}}\cdots\sum\limits_{p_{m}=1}^{s_{m}}\sum\limits_{q_{m}=1}^{t_{m}}
              \prod_{k=1}^{m}\textrm{Tr}(P_{k,(i,p_{k})}^{(j,q_{k})}\rho_{kl})\\
\leq&\sum\limits_{j=1}^{M}\sum\limits_{i=1}^{d}
 \sum\limits_{p_{1}=1}^{s_{1}}\sum\limits_{q_{1}=1}^{t_{1}}\cdots\sum\limits_{p_{m}=1}^{s_{m}}\sum\limits_{q_{m}=1}^{t_{m}}
              [\frac{1}{m}\sum\limits_{k=1}^{m}\big(\textrm{Tr}(P_{k,(i,p_{k})}^{(j,q_{k})}\rho_{kl})\big)^{2}]^{\frac{m}{2}}\\
\leq&\sum\limits_{j=1}^{M}\sum\limits_{i=1}^{d}
 \sum\limits_{p_{1}=1}^{s_{1}}\sum\limits_{q_{1}=1}^{t_{1}}\cdots\sum\limits_{p_{m}=1}^{s_{m}}\sum\limits_{q_{m}=1}^{t_{m}}
              [\frac{1}{m}\sum\limits_{k=1}^{m}\big(\textrm{Tr}(P_{k,(i,p_{k})}^{(j,q_{k})}\rho_{kl})\big)^{2}]\\
\leq&\frac{1}{m}\sum\limits_{i=1}^{m}\frac{\prod_{k=1}^{m} s_{k}t_{k}}{s_{i}t_{i}}\Big(\frac{M_{i}-1}{d_{i}}+\kappa_{i}\Big),
\end{array}
\end{equation}
where the inequality (\ref{M-1}) is used. It follows that
\begin{equation}
\begin{array}{rl}
J(\rho)=&\sum\limits_{j=1}^{M}\sum\limits_{i=1}^{d}
 \sum\limits_{p_{1}=1}^{s_{1}}\sum\limits_{q_{1}=1}^{t_{1}}\cdots\sum\limits_{p_{m}=1}^{s_{m}}\sum\limits_{q_{m}=1}^{t_{m}}
              \mathrm{Tr}\Big(\big(\otimes^{m}_{k=1}P_{k,(i,p_{k})}^{(j,q_{k})}\big) \rho \Big)\\
=&\sum\limits_{l}p_l \sum\limits_{j=1}^{M}\sum\limits_{i=1}^{d}
 \sum\limits_{p_{1}=1}^{s_{1}}\sum\limits_{q_{1}=1}^{t_{1}}\cdots\sum\limits_{p_{m}=1}^{s_{m}}\sum\limits_{q_{m}=1}^{t_{m}}
              \mathrm{Tr}\Big(\big(\otimes^{m}_{k=1}P_{k,(i,p_{k})}^{(j,q_{k})}\big) \rho_l \Big)\\
\leq&\frac{1}{m}\sum\limits_{i=1}^{m}\frac{\prod_{k=1}^{m} s_{k}t_{k}}{s_{i}t_{i}}\Big(\frac{M_{i}-1}{d_{i}}+\kappa_{i}\Big),
\end{array}
\end{equation}
i.e.  inequality (\ref{th4-1}) holds.

By the Cauchy-Schwarz inequality and the relation (\ref{M-1}), we deduce that
\begin{equation}
\begin{array}{rl}
J(\rho_l)
\leq&\sum\limits_{j=1}^{M}\sum\limits_{i=1}^{d}
 \sum\limits_{p_{1}=1}^{s_{1}}\sum\limits_{q_{1}=1}^{t_{1}}\cdots\sum\limits_{p_{m}=1}^{s_{m}}\sum\limits_{q_{m}=1}^{t_{m}}
              \textrm{Tr}(P_{a,(i,p_{a})}^{(j,q_{a})}\rho_{al})\textrm{Tr}(P_{b,(i,p_{b})}^{(j,q_{b})}\rho_{bl})\\
\leq&\sqrt{\sum\limits_{j=1}^{M}\sum\limits_{i=1}^{d}
 \sum\limits_{p_{1}=1}^{s_{1}}\sum\limits_{q_{1}=1}^{t_{1}}\cdots\sum\limits_{p_{m}=1}^{s_{m}}\sum\limits_{q_{m}=1}^{t_{m}}
              (\textrm{Tr}(P_{a,(i,p_{a})}^{(j,q_{a})}\rho_{al}))^{2}}
     \sqrt{\sum\limits_{j=1}^{M}\sum\limits_{i=1}^{d}
 \sum\limits_{p_{1}=1}^{s_{1}}\sum\limits_{q_{1}=1}^{t_{1}}\cdots\sum\limits_{p_{m}=1}^{s_{m}}\sum\limits_{q_{m}=1}^{t_{m}}
              (\textrm{Tr}(P_{b,(i,p_{b})}^{(j,q_{b})}\rho_{bl}))^{2}}\\
\leq& \prod_{k=1}^{m} s_{k}t_{k}
         \sqrt{\frac{1}{s_{a}t_{a}}(\frac{M_{a}-1}{d_{a}}+\kappa_{a})}
         \sqrt{\frac{1}{s_{b}t_{b}}(\frac{M_{b}-1}{d_{b}}+\kappa_{b})}.
\end{array}
\end{equation}
It implies that
\begin{equation}
\begin{array}{rl}
J(\rho)=&\sum\limits_{j=1}^{M}\sum\limits_{i=1}^{d}
 \sum\limits_{p_{1}=1}^{s_{1}}\sum\limits_{q_{1}=1}^{t_{1}}\cdots\sum\limits_{p_{m}=1}^{s_{m}}\sum\limits_{q_{m}=1}^{t_{m}}
              \textrm{Tr}[(\otimes^{m}_{i=1} P_{k,(i,p_{k})}^{(j,q_{k})})\rho]\\
=&\sum\limits_{l}p_l
 \sum\limits_{j=1}^{M}\sum\limits_{i=1}^{d}
 \sum\limits_{p_{1}=1}^{s_{1}}\sum\limits_{q_{1}=1}^{t_{1}}\cdots\sum\limits_{p_{m}=1}^{s_{m}}\sum\limits_{q_{m}=1}^{t_{m}}
              \textrm{Tr}[(\otimes^{m}_{i=1} P_{k,(i,p_{k})}^{(j,q_{k})})\rho_l]\\
\leq&\prod_{k=1}^{m} s_{k}t_{k}
         \sqrt{\frac{1}{s_{a}t_{a}}(\frac{M_{a}-1}{d_{a}}+\kappa_{a})}
         \sqrt{\frac{1}{s_{b}t_{b}}(\frac{M_{b}-1}{d_{b}}+\kappa_{b})},
\end{array}
\end{equation}
which completes the proof of inequality (\ref{th4-2}).
\hfill $\square$

\vspace{0.2cm} {\slshape Theorem 4$'$.}~Suppose that $\rho$ is a density matrix in $\mathbb{C}^{d_{1}}\otimes\mathbb{C}^{d_{2}}\otimes\cdots\otimes\mathbb{C}^{d_{m}}$, ~$\mathcal{D}_{k}=\{\mathcal{B}_{k,1},\mathcal{B}_{k,2},\cdots,\mathcal{B}_{k,M_k}\}$ be a set of MUBs on~$\mathbb{C}^{d_{k}}$, where ~$\mathcal{B}_{k,j}=\{|k_{i}^{j}\rangle\}_{i=1}^{d_k}$,
 and $d_{k}=s_{k}d+r_{1k}, M_{k}=t_{k}M+r_{2k}$,
where $d=\min \{d_{1},d_{2},\cdots,d_{m}\}$, $M=\min \{M_{1},M_{2},\cdots,M_{m}\}$.
Define
\begin{equation}\label{}
J(\rho)=
\sum\limits_{j=1}^{M}\sum\limits_{i=1}^{d}
\mathrm{Tr}\Big(\big(\otimes^{m}_{k=1}(\sum\limits_{u=1}^{s_{k}}\sum\limits_{v=1}^{t_{k}}
                                       |k_{i_{u}}^{j_{v}}\rangle\langle k_{i_{u}}^{j_{v}}|)\big) \rho\Big),
\end{equation}
then any fully separable state $\rho$ satisfies
\begin{equation}\label{th4'}
J(\rho)\leq \min_{1\leq a\neq b\leq m}\prod_{k=1}^{m} s_{k}t_{k}
\sqrt{\frac{1}{s_{a}t_{a}}(1+\frac{M_a-1}{d_a})}\sqrt{\frac{1}{s_{b}t_{b}}(1+\frac{M_b-1}{d_b})}.
\end{equation}

\vspace{0.2cm} {\slshape Theorem 4$''$.}~Suppose that $\rho$ is a density matrix in $\mathbb{C}^{d_{1}}\otimes\mathbb{C}^{d_{2}}\otimes\cdots\otimes\mathbb{C}^{d_{m}}$, $\mathcal{P}_{k}$ is a set of GSIC-POVMs on ~$\mathbb{C}^{k}$, with efficiency parameter ~$\alpha_{k}$, and $d'^2_k=s_k d^2_k+r_{k}$, where $d=\min \{d_{1},d_{2},\cdots,d_{m}\}$.
Define
\begin{equation}\label{}
J(\rho)=
\sum\limits_{n=1}^{d^2}
\mathrm{Tr}\Big(\big(\otimes^{m}_{k=1}(\sum\limits_{i=1}^{s_{k}}P_{k,(n,i)})\big) \rho\Big),
\end{equation}
then every fully separable state $\rho$ satisfies
\begin{equation}\label{th4''}
J(\rho)\leq \min_{1\leq a\neq b\leq m}\prod_{k=1}^{m} s_{k}
\sqrt{\frac{(\alpha_{a}d_a^{2}+1)}{s_{a}d_a(d_a+1)}}
\sqrt{\frac{(\alpha_{b}d_b^{2}+1)}{s_{b}d_b(d_b+1)}}.
\end{equation}

Using the above two bounds, not only multilevel multiparticle genuine entangled states, but also $k$-nonseparable states can be detected with the same method detailed discussed in Ref.\cite{1512.02853}.

\section{Conclusion and discussions}

Mutually unbiased measurements (MUMs) have been used to investigate entanglement detection and we obtained separability criteria for bipartite systems composed of a $d_1$-dimensional subsystem and a $d_2$-dimensional subsystem via sets of MUMs. The previous criteria are improved by taken into account more terms which were ignored before, leading to more effective bounds as we have proved. Moreover it should be noted that the method could be used to obtain some separable criteria via other measurements, such as MUBs and GSIC-POVMs as discussed. Noting the importance of the study on multiparty quantum entanglement, especially in higher-dimensional systems more than qubits, we have generalized our criteria to high dimensional multipartite systems presented in this paper, ameliorating the corresponding ones obtained previously \cite{1512.02853}.
These criteria is computationally simple and provide experimental implementation in detecting entanglement without full quantum state tomography, requiring only a few local measurements. It is worth noting that many other separability criteria may be improved with the method proposed in this paper.

\vspace{0.5cm} \noindent{\bf\large Acknowledgments }

\noindent
This work was supported by the National Natural Science Foundation
of China under Grant Nos: 11371005, 11475054; the Hebei Natural Science Foundation
of China under Grant No: A2016205145.
\end{document}